\documentclass[12pt]{article}
\usepackage{longtable}
\usepackage{epsfig,graphicx}
\usepackage{lscape}

\begin{document}

\begin{center}

\vspace*{1.0cm}

 {\Large \bf{Search for $\alpha$ decay of $^{151}$Eu to the first excited level of $^{147}$Pm using underground $\gamma$-ray
 spectrometry}}

 \vskip 1.0cm

{\bf F.A.~Danevich$^{a}$, E.~Andreotti$^{b}$, M.~Hult$^{b}$,
G.~Marissens$^{b}$, V.I.~Tretyak$^{a}$, A.~Yuksel$^{b}$}

\vskip 0.3cm

$^{a}${\it Institute for Nuclear Research, MSP 03680 Kyiv,
Ukraine}

$^{b}${\it  European Commission, Joint Research Centre, Institute
for Reference Materials and Measurements, Retieseweg, B-2440 Geel,
Belgium}

\end{center}

\vskip 0.5cm

\begin{abstract}{The alpha decay of $^{151}$Eu to the first excited level of
$^{147}$Pm ($J^\pi = 5/2^+$, $E_{exc}=91.1$ keV) was searched for
at the HADES underground laboratory ($\approx 500$ m w.e.). A
sample of high purity europium oxide with mass of 303 g and a
natural isotopic composition has been measured over 2232.8 h with
a high energy resolution ultra-low background n-type semi-planar
HPGe detector (40 cm$^3$) with sub-micron deadlayer. The new
improved half-life limit has been set as $T_{1/2} \geq 3.7\times
10^{18}$ yr at 68\% C.L. Possibilities to improve the sensitivity
of the experiment, which is already near the theoretical
predictions, are discussed. New half-life limit for $\alpha$ decay
of $^{153}$Eu is also set as $T_{1/2} \geq 5.5\times 10^{17}$ yr.}

\end{abstract}

\vskip 0.4cm


\vskip 0.4cm

\section{Introduction}

Natural europium consists of only two isotopes, $^{151}$Eu and
$^{153}$Eu, with the natural abundances of 47.81(0.06)\% and
52.19(0.06)\%, respectively \cite{Berglund:2011}. Both isotopes
are potentially $\alpha$ active with an $\alpha$ decay energy
$Q_{\alpha}=1964.9(1.1)$ keV and $Q_{\alpha}=272.5(2.0)$ keV,
respectively \cite{Audi:2003,Audi:2011}. The first indication an
$\alpha$ decay of $^{151}$Eu to the ground state of $^{147}$Pm
with the half-life $T_{1/2}=5^{+11}_{-3}\times 10^{18}$ yr was
obtained in \cite{Belli:2007a} with the help of a CaF$_2$(Eu) low
background scintillation detector.

Alpha decay of $^{151}$Eu is also energetically allowed to the
excited levels of $^{147}$Pm, with a highest probability of
transition to the first 91.1 keV $5/2^+$ level (the decay scheme
of $^{151}$Eu is shown in fig. 1). Theoretical estimations for
this $\alpha$ transition (obtained by using different approaches
\cite{Poenaru:1983,Buck:1991,Buck:1992,Denisov:2009}) are in the
range of $7\times 10^{18}-1\times 10^{20}$ yr, which is at the
level of present sensitivity accessible in low background
experiments. Indeed, recently the $\alpha$ decay of $^{190}$Pt to
the first excited level of $^{186}$Os with the half-life
$2.6\times10^{14}$ yr was observed by using an ultra-low
background HPGe detector and a sample of platinum with the natural
composition of isotopes \cite{Belli:2011}, despite very low
isotopic abundance of $^{190}$Pt (0.012\%). The decay
$^{151}$Eu$\to^{147}$Pm$^*$ seems to be a detectable process with
the present experimental technique, taking into account more than
three orders of magnitude higher isotopic abundance of $^{151}$Eu.

\nopagebreak
\begin{figure}[htbp]
\begin{center}
\mbox{\epsfig{figure=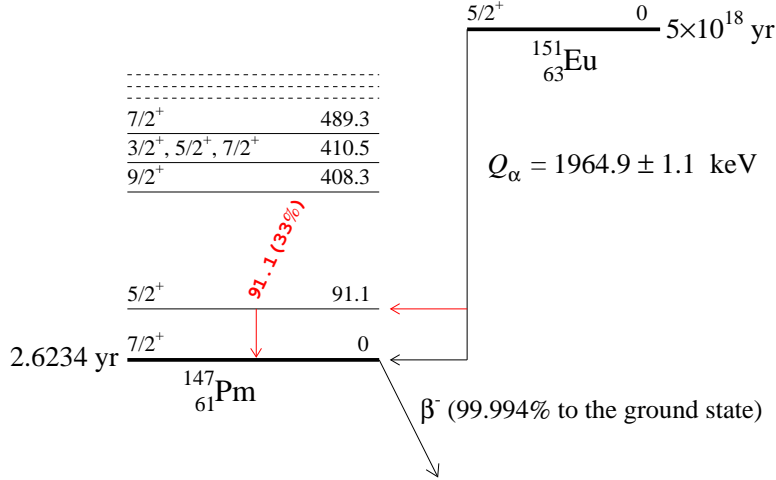,height=6.5cm}} \caption{(Color
online) Expected scheme of $\alpha$ decay of $^{151}$Eu (levels
above 489.3 keV are omitted). The energies of the levels and of
the de-excitation $\gamma$ quantum are given in keV
\cite{Firestone:1998,Nica:2009}.}.
\end{center}
\end{figure}

To our knowledge, there were two experimental attempts to detect
the process.
The limit $T_{1/2}\geq 2.4\times10^{16}$ yr was set in
the measurements with a HPGe detector with a small (2.72 g) sample
of Li$_6$Eu(BO$_3$)$_3$ crystal (containing only 1.1 g of Eu; the
purpose of the experiment was to investigate radioactive
contamination of the material) \cite{Belli:2007b}.
One order of magnitude higher limit $T_{1/2}\geq 6.0\times10^{17}$ yr was set
in the experiment with CaF$_2$(Eu) crystal scintillator \cite{Belli:2007a}.

One can also search for $\alpha$ decay of $^{153}$Eu to the ground
state of $^{149}$Pm by using a $\gamma$-ray detector because of
instability of the daughter $^{149}$Pm nucleus relative to
$\beta$ decay with half-life $T_{1/2}=53.08$ h and
$Q_{\beta}=1071$ keV (the decay scheme of $^{153}$Eu is shown in
fig. 2). Up to date only the limit $T_{1/2}\geq 1.1\times10^{16}$
yr on $\alpha$ decay of $^{153}$Eu was set in \cite{Belli:2007b}.

\nopagebreak
\begin{figure}[htbp]
\begin{center}
\mbox{\epsfig{figure=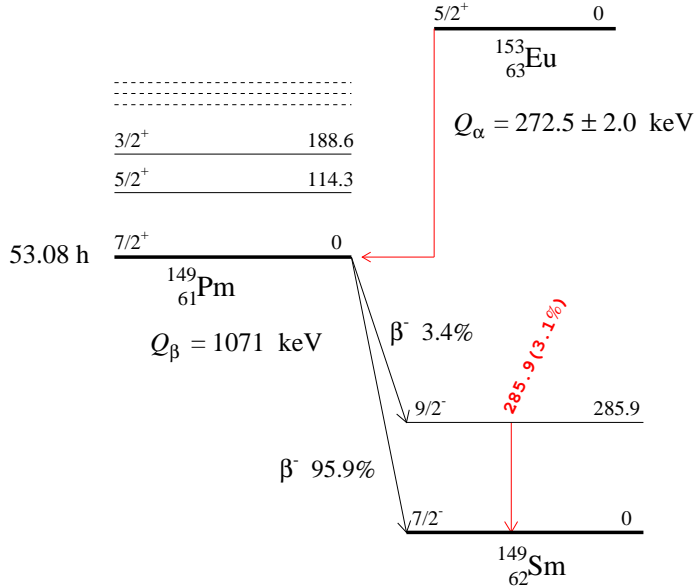,height=8.0cm}} \caption{(Color
online) Expected scheme of $\alpha$ decay of $^{153}$Eu. The
energies of the levels and of the deexcitation $\gamma$ quantum
are given in keV \cite{Firestone:1998,Singh:2004}. Only the main
branches of $^{149}$Pm $\beta$ decay are shown.}.
\end{center}
\end{figure}

The present paper describes a new search for $\alpha$ decay of $^{151}$Eu
to the first excited level of $^{147}$Pm and $\alpha$ decay of $^{153}$Eu to the ground state of $^{149}$Pm by using a high purity europium oxide sample installed on an ultra-low
background HPGe $\gamma$ detector.

\section{Materials and methods}

The measurements were performed in the HADES underground facility, which is located at the premises of the Belgian nuclear centre (SCK-CEN) and operated by EURIDICE
 \cite{HADES}. The depth of the laboratory is 225 m.
The sand and clay overburden of about 500 m of water equivalent
assures a muon flux reduction factor of about 5000.

A high purity europium oxide (Eu$_2$O$_3$) sample (in the form of
white powder) with mass 303 g, contained in a polyethylene bag,
was placed directly on the 0.7 mm thick aluminium endcap of the
HPGe-detector Ge-2 in HADES. The shape of the sample was almost
cylindrical (diameter 6 cm, height 11 cm) defined by the inner
copper shield. Detector Ge-2 is an ultra low-background n-type
semiplanar HPGe detector (volume 40 cm$^3$) with a sub-micron top
deadlayer. The shielding surrounding the detector consists of 15
cm of lead with an inner 10 cm layer of low radioactive lead (the
activity of $^{210}$Pb $<0.5$ Bq/kg) originating from old French
monuments. There is also an inner lining of 6 cm of freshly
produced electrolytic copper. The $^{222}$Rn activity in air of
the laboratory is rather low (average of about 6 Bq/m$^3$) and is
reduced further inside the shield by flushing it with the boil-off
nitrogen from the detector's Dewar. Although detector Ge-2 is the
smallest detector in HADES, it was chosen because of its excellent
energy resolution and because the main interest of this study is
the energy region around 91 keV. The detector energy resolution
(the full width at the half of maximum, FWHM, keV) can be
approximated in the energy region of $60-662$ keV by function
FWHM~$=0.4699+0.02928\sqrt{E_{\gamma}}$, where $E_{\gamma}$ is
energy of $\gamma$ quanta in keV. In particular, FWHM~$=0.749$ keV
at 91 keV.

The stability of the measurement system is very high due to the
underground location with a cast iron lining and due to the use of
a UPS (uninterruptable power supply). Generally, one spectrum per
day is collected and acquisition is stopped when liquid nitrogen
is filled. The energy stability of the system was possible to
check daily due to the two main peaks at 122 keV and 344 keV of
$^{152}$Eu and with a bin-width of 82 eV they showed no detectable
drift. For long measurements, the first spectrum after changing a
sample is usually excluded in order to assure that radon is
removed by the boil-off nitrogen.

The data with the Eu$_2$O$_3$ sample were collected over 93.03
days, while the background spectrum of the detector was measured
over 68.95 days. The energy spectrum accumulated with the
Eu$_2$O$_3$ sample in the energy range of $10-670$ keV is shown in
fig. 3 together with the background data (normalized to the
measurement time of the sample).

\nopagebreak
\begin{figure*}[htbp]
\begin{center}
\mbox{\epsfig{figure=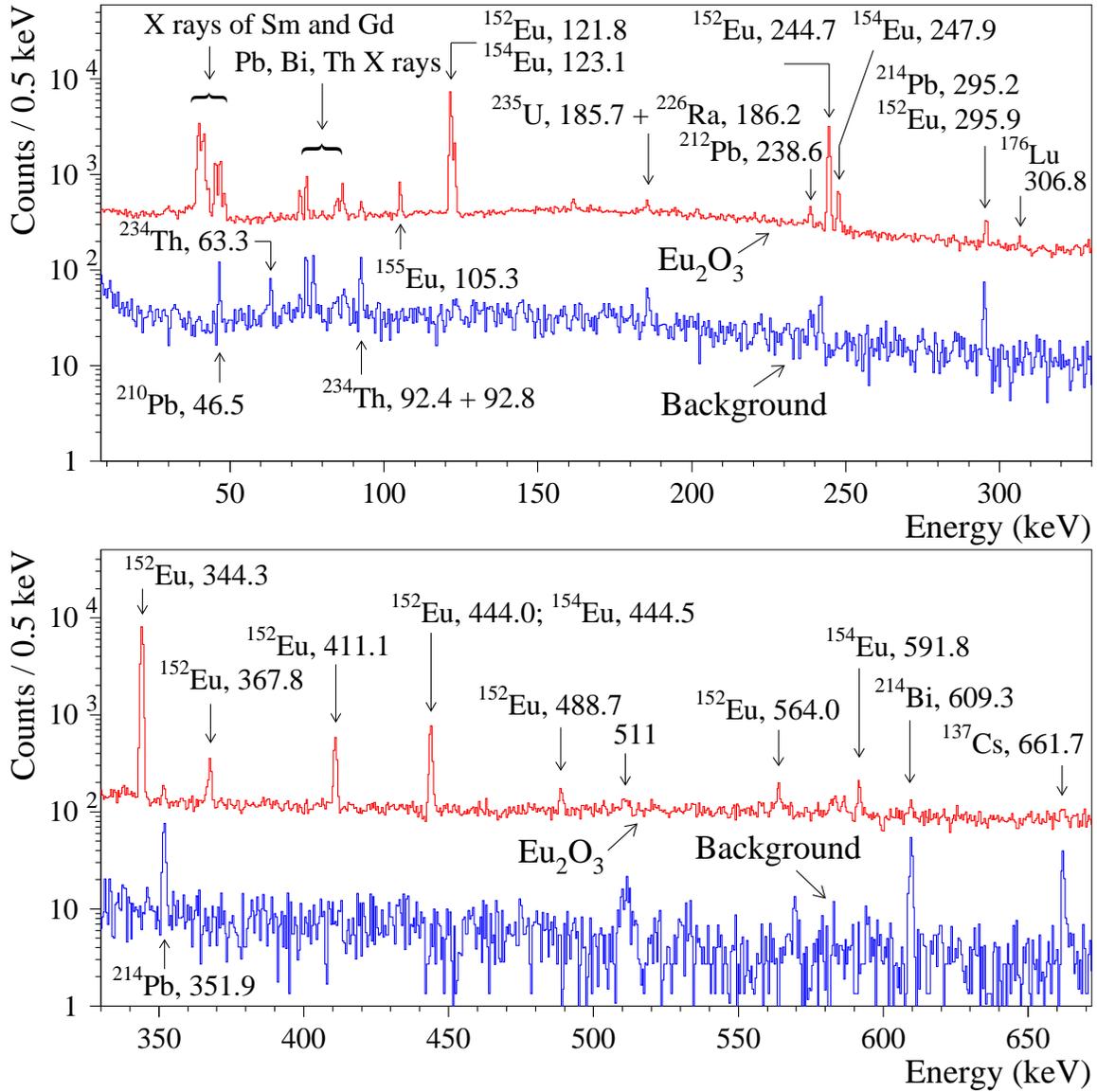,height=16.0cm}} \caption{(Color
online) Energy spectrum of the Eu$_2$O$_3$ sample measured during
2232.8 h in the $10-330$ keV energy interval (upper part), and in
the $330-670$ keV energy interval (lower part). The background
spectrum (measured during 1654.7 h, normalized to 2232.8 h) is
also shown. The energies of $\gamma$ lines are in keV
\cite{Firestone:1998}.}
\end{center}
\end{figure*}

With the aim of quantifying radionuclides emitting gamma-rays up
to 2.7 MeV, an additional measurement for 15.4 days was carried
out using detector Ge-5. This is a planar HPGe-detector from
Canberra of a so-called BEGe-type (Broad Energy Germanium). It is
a p-type crystal with a submicron deadlayer. Its diameter and
thickness are 80 mm and 30 mm, respectively, resulting in a
relative efficiency of 50\% (volume 150 cm$^3$). The endcap of the
detector is fabricated from high purity aluminium. The Eu$_2$O$_3$
sample of a cylindrical shape (diameter 8.1 cm, height 6.2 cm) was
placed directly on the endcap of the HPGe-detector Ge-5.

\section{Results}

\subsection{Radioactive contamination of Eu$_2$O$_3$ sample}

The comparison of the Eu$_2$O$_3$ and of the background spectra
shows that the europium sample is contaminated by daughters of
$^{232}$Th and $^{238}$U, and radioactive europium isotopes
$^{152}$Eu, $^{154}$Eu and $^{155}$Eu. There are also peaks in the
spectrum with the energies $201.7\pm0.2$ keV and $306.5\pm0.2$
keV, which indicate presence of $^{176}$Lu in the sample.
Radioactive $^{152}$Eu and $^{154}$Eu nuclei were produced by
neutron captures in $^{151}$Eu and $^{153}$Eu, respectively.
$^{155}$Eu can be produced by the double neutron capture reaction
$^{153}$Eu(n,$\gamma)\to^{154}$Eu(n,$\gamma)\to^{155}$Eu and also
due to fission of uranium and thorium. Indeed, europium is
extracted from minerals with typically high concentration of
thorium and uranium: bastnaesite, loparite, xenotime, and
monazite. There are no unidentified peaks in the spectrum.

Massic activities of the contaminant radionuclides ($A$) were calculated
with the formula:
\begin{center}
$$A =
(S_{sample}/t_{sample}-S_{bg}/t_{bg})/(\vartheta\cdot\varepsilon\cdot
m),$$
\end{center}
where $S_{sample}~(S_{bg})$ is the area of a peak in the sample
(background); $t_{sample}~(t_{bg})$ is the time of the sample
(background) measurement; $\vartheta$ is the $\gamma$-ray emission
probability \cite{Firestone:1998}, $\varepsilon$ is the efficiency
of the full energy peak detection; $m$ is the mass of the sample.
The full energy peak efficiencies were calculated using the Monte
Carlo code EGS4 \cite{EGS4}. A detailed model of the detector,
sample and shield were used for the calculations. Coincidence
summing corrections for cascading gamma-rays were also included in
the simulation. The systematic uncertainty of the calculated
efficiency is at the level of 5\% as confirmed by several
proficiency testing exercises. The systematic relative uncertainty
of the detection efficiency calculation arising from the non
perfect cylindrical geometry of the sample (contained in a plastic
bag) was estimated to be $\approx 10\%$. Estimations of
radioactive contamination of the sample from the 15.4-day
measurement on Ge-5 were performed in the same way. The summary of
radioactive contamination of the Eu$_2$O$_3$ sample is presented
in table 1. The results of the independent measurements are in
reasonable agreement, taking into account a systematic uncertainty
due to the non perfect cylinder shapes of the sample in both the
measurements.

\nopagebreak
\begin{table}[htb]
\caption{Radioactive contamination of the Eu$_2$O$_3$ sample
measured in HADES by the HPGe $\gamma$ detectors Ge-2 and Ge-5. All uncertainties are presented as the combined standard uncertainty (k=1). The reference dates for the activities are October 2011 for Ge-2 and April 2012 for Ge-5.}
\begin{center}
\begin{tabular}{|l|l|l|l|}

 \hline
  Decay     & Radionuclide  & \multicolumn{2}{c|}{Massic activity (mBq/kg)} \\
  chain     & ~             & \multicolumn{2}{c|}{~} \\
 \cline{3-4}
  ~         & ~             & Ge-2      &  Ge-5 \\
 \hline
 ~          & $^{40}$K          & -         & $420\pm40$ \\
 ~          & $^{60}$Co         & -         & $\leq 9$ \\
 ~          & $^{137}$Cs        & $\leq 3$      & $\leq 18$ \\
 ~          & $^{138}$La        & -             & $\leq 7$ \\
 ~          & $^{152}$Eu        & $2520\pm280$  & $3060\pm330$ \\
 ~          & $^{154}$Eu        & $380\pm40$    & $440\pm50$ \\
 ~          & $^{155}$Eu        & $210\pm40$    & $290\pm40$ \\
 ~          & $^{176}$Lu        & $4\pm2$       & $7\pm2$ \\
 \hline
 $^{232}$Th & $^{228}$Ra        & $23\pm7$      & $\leq 20$\\
 ~          & $^{228}$Th        & $23\pm5$      & $23\pm8$ \\
 \hline
 $^{235}$U  & $^{235}$U         & $10\pm3$      & $10\pm3$\\
 ~          & $^{231}$Pa        & $\leq100$     & $\leq210$\\
 ~          & $^{227}$Ac        & $\leq 16$     & $\leq100$ \\
 \hline
 $^{238}$U  & $^{234m}$Pa       & $290\pm130$   & $\leq910$\\
 ~          & $^{226}$Ra        & $\leq 15$     & $17\pm4$\\
 ~          & $^{210}$Pb        & $\leq 1200$   & - \\

\hline
\end{tabular}
\end{center}
\end{table}

\subsection{Limit on $\alpha$ decay of $^{151}$Eu to the first excited level of $^{147}$Pm}

Gamma quanta with an energy of 91.1 keV should be emitted after alpha
decay of $^{151}$Eu to the first excited level of $^{147}$Pm.
There is no peak at energy 91 keV in the energy spectrum
accumulated with the Eu$_2$O$_3$ sample (see fig. 4). Therefore we
can only set a lower half-life limit ($\lim T_{1/2}$) on the effect
according to the formula:

\begin{center}
$\lim T_{1/2} = N \cdot \varepsilon \cdot \vartheta \cdot t \cdot \ln 2 /
\lim S$,
\end{center}
where $N$ is the number of $^{151}$Eu nuclei
($4.96\times10^{23}$), $\varepsilon$ is the detection efficiency,
$\vartheta$ is the $\gamma$ yield ($\vartheta=0.33$ for the level
91.1 keV due to a high electron conversion coefficient of 2.03
\cite{Nica:2009}), $t$ is the measuring time, and $\lim S$ is the
number of events of the effect searched for which can be excluded
at a given confidence level (C.L.). The detection efficiency for
91.1 keV $\gamma$ quanta emitted in the Eu$_2$O$_3$ sample was
calculated with the EGS4 package as $\varepsilon=0.00434$ (see
sect. 3.1).

\nopagebreak
\begin{figure}[htbp]
\begin{center}
\mbox{\epsfig{figure=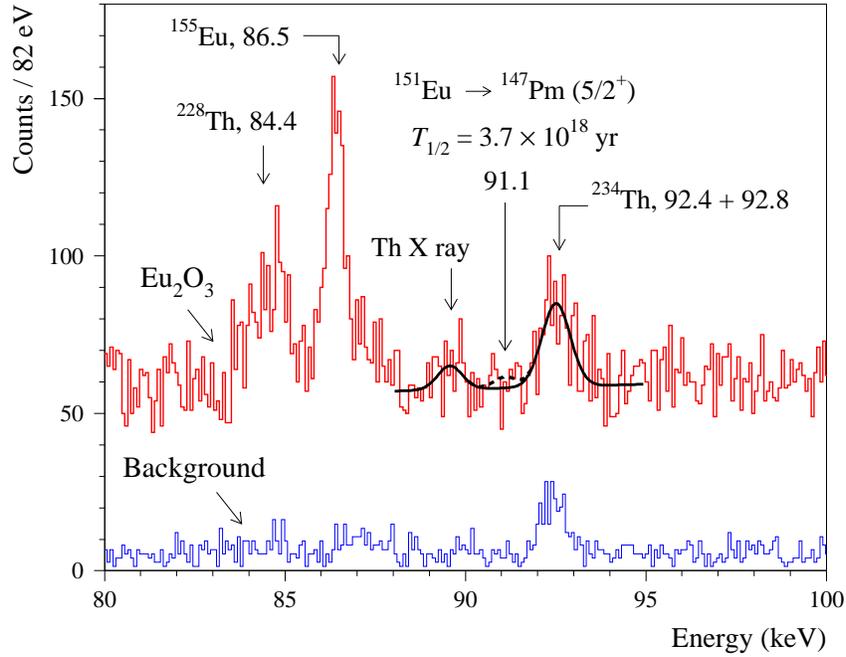,height=9.5cm}} \caption{(Color
online) The energy spectrum measured over 2232.8 h with the
Eu$_2$O$_3$ sample in the $80-100$ keV energy interval. Fit of the
spectrum in the vicinity of 91.1 keV $\gamma$ peak is shown by
solid line. The expected 91.1 keV $\gamma$ peak from $\alpha$
transition $^{151}$Eu $\to$ $^{147}$Pm$^*$ corresponds to the
half-life $T_{1/2}=3.7\times10^{18}$ yr excluded at 68\% C.L.;
shown by the dashed line. The background spectrum (measured during
1654.7 h, normalized here to 2232.8 h) is also presented. The
energies of $\gamma$ lines are in keV.}
\end{center}
\end{figure}

To estimate a value of $\lim S$ for
the 91.1 keV peak, the energy spectrum
accumulated with the Eu$_2$O$_3$ sample was fitted in the energy
region $88-95$ keV. The model to fit the data was constructed from
a Gaussian function with centre at the energy of 91.1 keV and the energy
resolution FWHM = 0.749 keV (the $\gamma$ peak searched for), a
linear function which describes the background, and two Gaussians to
take into account the neighbouring peaks with the energies
$\approx90$ keV (Th X ray) and $\approx92.6$ keV (92.4 and 92.8 keV peaks
of $^{234}$Th from the $^{238}$U chain).

A fit by the chisquare method ($\chi^{2}/$n.d.f.$~= 89/78=1.14$,
where n.d.f. is number of degrees of freedom) results in the area
of the peak searched for $S=7\pm27$ counts, which gives no
evidence for the effect. In accordance with the Feldman-Cousins
procedure \cite{Feldman:1998}, we took $\lim S=34$ counts which
can be excluded at 68\% C.L. Thus we obtained the following
limit on $\alpha$ decay of $^{151}$Eu to the first $5/2^+$ 91.1
keV excited level of $^{147}$Pm:

\begin{center}
$T_{1/2}^{\alpha}[^{151}$Eu$~\rightarrow~^{147}$Pm~$(5/2^{+},~91.1~$keV$)]\geq~3.7\times10^{18}$
yr.
\end{center}

The data collected with the Eu$_2$O$_3$ sample also allow
searching for all possible $\alpha$ decays of $^{153}$Eu to
$^{149}$Pm (including the most probable decay to the ground state)
due to $\beta$ instability of the daughter isotope (see fig. 2).
The most intense $\gamma$ line of $^{149}$Pm has an energy of 285.9
keV and a yield $\vartheta=0.031$ \cite{Singh:2004}. The detection
efficiency of the $\gamma$ quanta is $\varepsilon=0.0114$. Part of
the spectrum in the energy interval $278–-298$ keV is shown in
fig. 5. There is no peculiarity in the spectrum accumulated with
the europium oxide sample, which can be ascribed to the 285.9 keV
$\gamma$ peak. However, there are two peculiarities (at 284.3 keV
and 284.9 keV, see fig. 5) immediately to the left of the sought
285.9 keV peak with a somewhat higher number of counts compared to
the surrounding channels. Calculations show that these
"structures" are not peaks as they fall below the detection limit.
We attribute them to counting statistical variations. A fit in the
energy interval $282-290$ keV by a model, consisting of the
Gaussian function to describe the peak searched for plus
polynomial function of the second degree to describe the
background, gives an area of the effect searched for as $34\pm27$
events, which is no evidence for the effect (see fig. 5). Taking
$\lim S=61$ we set the following half-life limit on $\alpha$
decay of $^{153}$Eu to the ground state of $^{149}$Pm at 68\%
C.L.:

\begin{center}
$T_{1/2}^{\alpha}(^{153}$Eu)~$\geq~5.5\times10^{17}$ yr.
\end{center}

\nopagebreak
\begin{figure}[htbp]
\begin{center}
\mbox{\epsfig{figure=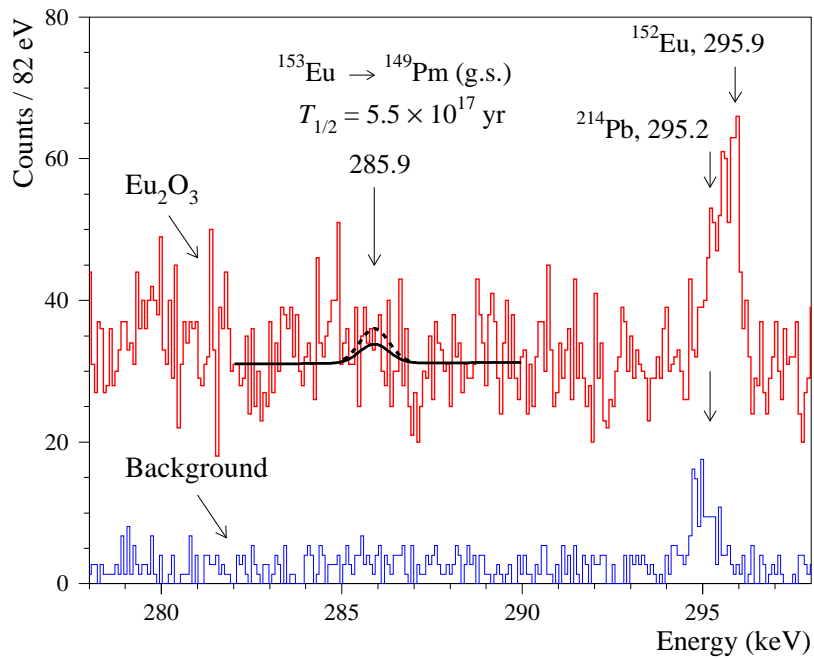,height=9.5cm}} \caption{(Color
online) Fragment of the energy spectra accumulated over 2232.8 h
with the Eu$_2$O$_3$ sample. Fit of the spectrum in the energy
interval $282-290$ keV is shown by solid line. The expected 285.9
keV $\gamma$ peak from $\beta$ decay of $^{149}$Pm (which could
appear after $\alpha$ decay of $^{153}$Eu) is shown by the dashed
line. Area of the peak corresponds to the half-life of $^{153}$Eu
$T_{1/2}=5.5\times10^{17}$ yr excluded at 68\% C.L. The background
spectrum (normalized to the time of measurements with the sample)
is also presented.}
\end{center}
\end{figure}

A summary of the obtained results in comparison with the previous
studies is given in table 2.

\begin{table*}[htbp]
\caption{Half-life limits on $\alpha$ decay of Eu isotopes (given
at 68\% C.L.; the half-life limits in work \cite{Belli:2007b} are
given at 90\% C.L.) and comparison with the theoretical
predictions. The values in columns $3-5$ are presented in years.}
\begin{center}
\begin{tabular}{|lllll|}
\hline
 Alpha                           & Level of daughter     & \multicolumn{2}{c}{Experimental $T_{1/2}$}                             & Theoretical \\
 transition,                     & nucleus               & Previous                                      & This work              & estimations  \\
 $Q_{\alpha}$ \cite{Audi:2003,Audi:2011} & ~             & result                                        & ~                      & ~ \\
 \hline
 ~ & ~ & ~ & ~ & ~ \\
 $^{151}$Eu $\to$ $^{147}$Pm     & $5/2^+$, 91.1 keV     & $\geq 2.4\times10^{16}$ \cite{Belli:2007b}    & $\geq3.7\times10^{18}$ & $9.7\times10^{19}$$^{~a}$ \\
 1964.9(1.1) keV                 & ~                     & $\geq 6.0\times10^{17}$ \cite{Belli:2007a}    & ~                      & $7.4\times10^{18}$$^{~b}$ \\
 ~                               & ~                     & ~                                             & ~                      & $4.5\times10^{19}$$^{~c}$ \\
 ~ & ~ & ~ & ~ & ~ \\
 $^{153}$Eu $\to$ $^{149}$Pm     & $7/2^+$, 0 keV (g.s.) & $\geq 1.1\times10^{16}$ \cite{Belli:2007b}    & $\geq5.5\times10^{17}$ & $5.2\times10^{144}$$^{~a}$ \\
 272.5(2.0) keV                  & ~                     & ~                                             & ~                      & $2.8\times10^{141}$$^{~b}$ \\
 ~                               & ~                     & ~                                             & ~                      & ~ \\
 \hline
 \multicolumn{5}{l}{$^{a}$ Calculated with semiempirical formulae
 \cite{Poenaru:1983}.}\\
 \multicolumn{5}{l}{$^{b}$ Calculated with the cluster model of ref. \cite{Buck:1991,Buck:1992}.}\\
 \multicolumn{5}{l}{$^{c}$ Calculated with the approach of ref. \cite{Denisov:2009}.}\\
\end{tabular}
\end{center}
\label{tb:summary}
\end{table*}

\section{Discussion}

Half-life values for the $\alpha$ decays of the Eu isotopes were
calculated here with the cluster model of ref.
\cite{Buck:1991,Buck:1992} and with semiempirical formulae
\cite{Poenaru:1983} based on liquid drop model and description of
$\alpha$ decay as a very asymmetric fission process. The
approaches were tested with a set of experimental half-lives of a
few hundred $\alpha$ emitters and demonstrated good agreement
between calculated and experimental $T_{1/2}$ values, mainly
inside the factor of $2-3$. We also successfully used these works
to predict $T_{1/2}$ values of the long-living $\alpha$ active
$^{180}$W \cite{Danevich:2003} and $^{151}$Eu \cite{Belli:2007a}
obtaining adequate agreement between the first experimentally
measured and calculated results. The results of the theoretical
predictions for the $\alpha$ decays of europium isotopes
considered in the present study are presented in table 2, together
with result of the recent work \cite{Denisov:2009}. One can see
that our experimental limit for the $\alpha$ decay $^{151}$Eu
$\to$ $^{147}$Pm$^*$ is not so far from the theoretical
predictions.

The sensitivity of the experiment can be improved by purification
of the Eu$_2$O$_3$ sample from potassium, uranium and thorium.
Nevertheless, the main source of the background in the experiment is
contamination of the sample by radioactive europium isotopes
$^{152}$Eu, $^{154}$Eu and $^{155}$Eu. One could overcome this
problem by using europium extracted from minerals with low
concentration of uranium and thorium to avoid production of radioactive
Eu nuclides. In this case sensitivity of an experiment can
be improved by one order of magnitude (also assuming $2-3$ times longer
measurements and optimization of efficiency).
Further improvement of sensitivity could be achieved
by using a Li$_6$Eu(BO$_3)_3$ crystal \cite{Belli:2007b} as
a cryogenic scintillating bolometer with typically very high
energy resolution for alpha particles (at the level of a few keV),
high detection efficiency, and excellent particle discrimination
ability which allow to distinguish single $\alpha$ events, $\alpha$
signals with admixture of $\gamma$ quanta, and pure
$\gamma(\beta)$ events. This approach was recently successfully
applied in work \cite{Beeman:2012} to detect $\alpha$ transitions
of $^{209}$Bi to the ground state and the first excited level
of $^{205}$Tl with the half-life
$(2.01\pm0.08)\times 10^{19}$ yr.

\section{Summary}

The alpha decays of naturally occurring europium isotopes, which
are accompanied by emission of $\gamma$ quanta, were searched for
with the help of the ultra-low background HPGe detector in the
HADES underground laboratory of the Institute for Reference
Materials and Measurements (Geel, Belgium). The new improved
half-life limit for alpha decay of $^{151}$Eu to the first excited
level of $^{147}$Pm ($J^\pi = 5/2^+$, $E_{exc}=91.1$ keV) is set
as: $T_{1/2}\geq 3.7\times10^{18}$ yr. This value is not so far
from the theoretical predictions, which are in the range of
$7\times 10^{18}-1\times 10^{20}$ yr.

The sensitivity of the experiment can be improved by one order of
magnitude by using a radiopure europium sample, increase of the
exposure and optimization of efficiency.
Taking into account the theoretical estimations, such an
improvement of sensitivity could lead to detection of the alpha
decay of $^{151}$Eu to the first excited level of $^{147}$Pm with
the half-life on the level of $10^{19}$ yr. Further improvement of
the sensitivity could be achieved by using Li$_6$Eu(BO$_3$)$_3$
crystals as scintillating bolometers, as proposed in
\cite{Belli:2007b}.

As a by-product of the experiment, the new improved $T_{1/2}$
limit for $\alpha$ decay of $^{153}$Eu to the ground state of
$^{149}$Pm was set as $T_{1/2} \geq 5.5\times10^{17}$ yr. However,
due to the very small energy release in this $\alpha$ decay, the
limit is still many orders of magnitude far from the theoretical
expectations.

\section{Acknowledgements}

The work performed by the HADES staff of Euridice is gratefully
acknowledged. F.A.~Danevich and V.I.~Tretyak were supported in
part through the Space Research Program of the National Academy of
Sciences of Ukraine. We are grateful to V.Yu.~Denisov for the
theoretical half-life for the $\alpha$ decay $^{151}$Eu $\to$
$^{147}$Pm$^*$ (91.1 keV) absent in \cite{Denisov:2009} but kindly
calculated on our request.

\end{document}